\documentclass[10pt]{article}
\usepackage{graphicx} 
\begin{document}
\begin{center}
{\bf ON THE ONE-LOOP CORRECTION OF  $\varphi^4$ THEORY IN HIGHER DIMENSIONS
}
\\[.6cm]
Rizwan Ul Haq Ansari\footnote{ph03ph14@uohyd.ernet.in } and P.K. Suresh\footnote{ pkssp@uohyd.ernet.in}\\[.4cm]

{ \it School of Physics,University of Hyderabad,\\
Hyderabad-500046.India.
}
\end{center}

\begin{abstract}
We have considered  $ \varphi^4$ theory in higher dimensions. Using functional diagrammatic approach, we computed the one-loop correction to effective potential of the scalar field in five dimensions. It is shown that $ \varphi^4$ theory can be regularised in five dimensions. Temperature dependent one-loop correction and critical temperature $\beta_c$ are computed and  $\beta_c$ depends on the fundamental scale $\cal{M}$ of the theory. A brief discussion of symmetry restoration  is also presented. The nature of phase transitions is examined and is of second order.
\end{abstract}



\section{Introduction}
Effective potential plays a very important role in  quantum field theory and much have been studied about it for various types of fields. It is very useful in the study of symmetry breaking, since the broken symmetry of classical potential may be restored due to quantum corrections. The aim of the present paper  is to study the behaviour of effective potential of a scalar field in higher dimensions. The motivation for this study is  due to recent developments in higher dimensional theories which try to bring gravity and other fundamental interactions in single theory \cite{1,2}
. In recent times scalar fields have been studied in higher dimensions to address many issues in particle physics and cosmology like inflation, cosmological constant, dark energy, etc.

In this paper, we study  the effective potential formally by calculating the one-loop approximation and critical temperature for symmetry restoration in higher dimensions. It is  interesting to see how the  one-loop correction and  critical temperature will behave in higher dimensions. We consider $\varphi^4$ theory in five  dimensions, but it is  known $\varphi^4$ theory is not renormalisable in five dimensions. Non-renormalisibility of the theory  does not mean that theory is not interesting, it may be part of high energy theory which is renormalisable. Another feature of the scalar field in five dimensions is that  the coupling constant  $\lambda$ is dimensionful. To calculate effective potential, we follow   the diagrammatic approach up to one-loop order. The momentum integral is divergent in the expression for one-loop correction and we apply a cut-off $\Lambda $ to compute it. In the present work we are able to get a finite expression for one-loop effective potential and show that it is possible to regularize  $\varphi^4$ theory in five dimensions. The  critical temperature $\beta_c$ can be found by using functional diagrmatic  approach of Dolan and Jackiw \cite{3}.

The paper is organised as follows in  section II we compute the zero temperature one-loop approximation. Section III   deals with temperature one loop correction and the critical temperature. Results and conclusions are given in section IV.
\section{Effective  Potential and One-loop Correction}
 Consider  a scalar $\varphi^4$ theory in five dimensions (5D), the signature of the metric is (+,-,-,-,-). The action for the scalar field with a source term can be written as
\begin{eqnarray}
S^{J}&=& \int d^{5}x[\frac{1}{2}\partial^{\mu}\varphi
\partial_{\mu}\varphi-\frac{m^{2}}{2}\varphi^{2}-\frac{\lambda}{4!\cal{M}}\varphi^{4}+J\varphi].
\end{eqnarray} \\
 The tree level potential is given by
\begin{equation}
  V^{0}=\frac{1}{2} m^{2} \varphi^{2}+\frac{\lambda}{4!\cal{M}}\varphi^{4}-J \varphi .
 \end{equation}
Since the coupling constant $\lambda$ acquires dimensions in 5D, it is divided by $\cal{M}$ to make it dimensionless, which is the fundamental scale of the theory. 

The generating functional  is
\begin{equation}
  Z[J]=e^{\frac{i}{\hbar}W[J]}.
\end{equation}
 Where W[J] is the generating functional for the connected Green's functions   \cite{4}.
In quantum field theory we study effective action $\Gamma$ which is the Legendre transform of W[J]. In 5D it can be written as
\begin{eqnarray}
\Gamma[\hat{\varphi}_c]&=&W[J]-\int d^{5}x J(x) \varphi_{c}(x).
\end{eqnarray}
Where
\begin{equation}
 \hat{\varphi}_{c}= \frac{\delta W[J]}{\delta J(x)}
 \end{equation}
and
 \begin{equation}
  \frac{\delta \Gamma}{\delta \hat{\varphi}_{c}(x)} \vert_{\hat{\varphi}_{c}  \neq 0}=0 .
\end{equation}
Alternatively expanding the effective action in powers of momentum
and neglecting higher order terms in $\hat{\varphi}_{c}$.
 And in limit  $J \rightarrow 0 $, for \ $\hat{\varphi} \rightarrow \varphi_{c}$ is constant,
 we get
 \begin{equation}
 \Gamma[\varphi_c]=-(2\pi)^{5} \delta(0)V(\varphi_c)=-\Omega V(\varphi_c).
 \end{equation}
 Where $V(\varphi)$ is effective potential in five dimensions and $\Omega$ is  the space-time volume.

Now the  effective potential contains quantum corrections of different orders. We can calculate these corrections   peturbatively by loop expansion. Hence
\begin{equation}
V=V^{0}+\hbar V^1 +O(\hbar^2),
\end{equation}
where $V^{0}$ is the tree potential and  $V^1$ is the one-loop
correction. The one-loop contribution is computed using the
techniques developed earlier and can be written in closed form. The
one loop contribution to effective potential in 5D is obtained as
\begin{equation}
V^{1}(\varphi_c)= -\frac{i\hbar}{2} \int \frac{d^{5}k}{(2\pi)^{5}} \ln(-k^2 + M^2 ),
\end{equation}
 where shifted mass
$ M^{2}=m^2+ \frac{\lambda}{2\cal{M}} \varphi_c^2 $.\\

 Now rotating to Euclidean space, $k_0 =ik_E$
 \begin{eqnarray}
V^{1}(\varphi_c)&= &\frac{\hbar}{2} \int \frac{d^{5}k_E}{(2\pi)^{5}} \ln(k_{E}^2 + M^2 )
\\
 &= &\frac{\hbar}{24\pi^3} \int  k^4_E dk_E \ln(k_{E}^2 + M^2 ).
\end{eqnarray}
But the integral in eqn (11) is divergent. It has ultra-violet
divergence and its degree of divergence is three. To make the theory
meaningful we should make it finite by using renormalisation
technique. For this we first regularise the theory by applying
cut-off regularisation method. So, we introduce cut off $\Lambda$ on
all loop momenta and solve the integral to get
 \begin{eqnarray}
\nonumber V^{1}(\varphi_c)& =& \frac{\hbar}{24\pi^3} [ \frac{-2
\Lambda^5}{25}+ \frac{2}{15} \Lambda^3 M^2  -\frac{6}{15}
\Lambda M^4
+\frac{3}{15}iM^5\ln(\frac{M-i\Lambda}{M})\nonumber
\\
 &&-\frac{3}{15}iM^5\ln(\frac{M+i\Lambda}{M})
+\frac{3}{15}\Lambda^5 \ln(M^2+\Lambda^2) ] .
\end{eqnarray}
Thus an expression for one-loop potential in five dimensions is
obtained. But the expression is still  divergent when $\Lambda
\rightarrow \infty$. In general this infinities are  absorbed by
adding counter terms to the above expression and using suitable
renormalisation conditions. Since the $\varphi^4$ theory is not
renormalisable in 5 dimensions  it is interesting to see if we can
add counterterms to the Lagrangian and get rid of atleast some of
the divergent amplitude. The effective potential with counter terms is
 \begin{eqnarray}
\nonumber V& =& \frac{m^{2}+ \delta m^{2}}{2}
\varphi_c^{2}+\frac{\lambda + \delta\lambda}{4!\cal{M}}\varphi_c^{4} +
\frac{\hbar}{24\pi^3} [ \frac{-2 \Lambda^5}{25}+ \frac{2}{15}
\Lambda^3 M^2 -\frac{6}{15} \Lambda M^4 \\
&& + \frac{3}{15}iM^5\ln(\frac{M-i\Lambda}{M})
-\frac{3}{15}i M^5\ln(\frac{M+i\Lambda}{M})
+\frac{3}{15}\Lambda^5 \ln(M^2+\Lambda^2)]
\end{eqnarray}
where $\delta m^{2}$  and $\delta\lambda$ are  mass and coupling
constant renormalization counterterms. We use the following renormalisation conditions
\begin{eqnarray}
\frac{d^{2}V}{d\varphi_c^2} \vert_{\varphi_c=0} =m^2 ,\\
\frac{d^{4}V}{d\varphi_c^4} \vert_{\varphi_c=0} = \frac{\lambda}{\cal{M}}.
\end{eqnarray}
After detailed calculations, imposing conditions  (14) and (15), we obtain the counter terms as
\begin{equation}
\delta m^{2} = \frac{1}{3} \Lambda^3 \, \lambda + \Lambda
\,\lambda m^2 \,+\frac{5 \pi}{10}\lambda m^3 + \frac{6 \lambda}{5
\Lambda} m^4,
\end{equation}
and
\begin{equation}
\delta\lambda = 3\Lambda \lambda^2 + \frac{45}{20} \pi \lambda^2
\Lambda + \frac{36}{5\Lambda}m^2 \lambda^2.
\end{equation}
Now substiuting this counter terms in eqn (13), we get the effective potential
\begin{eqnarray}
V&=&\frac{1}{2} m^{2}
\varphi_c^{2}+\frac{\lambda}{4!\cal{M}}\varphi_c^{4}+ \frac{\hbar}{24
\pi^3} [-\frac{2}{25} \Lambda^5 + \frac{3}{15} \Lambda^3 m^2
-\frac{13}{30} \Lambda m^4 -\frac{\pi}{5}M^5
\nonumber \\
 && - \frac{2}{5 \Lambda}(m^6+ \frac{\lambda^3}{8 {\cal{M}}^3}\varphi_c^6)+ \frac{1}{2}\varphi_c^2(\frac{5\pi \lambda}{10\cal{M}} m^3)+ \frac{1}{4!}\varphi_c^4 \frac{45 \pi \lambda^2}{20 {\cal{M}}^2} m].
\end{eqnarray}
 Neglecting the field independent terms, and terms which vanish
in the limit $\Lambda\rightarrow \infty$ in eqn (18), the  effective potential
with one loop correction is obtained as
\begin{eqnarray}
V&=&\frac{1}{2} m^{2} \varphi_c^{2}+\frac{\lambda}{4!\cal{M}}\varphi_c^{4}+\frac{\hbar}{24 \pi^3} [ -\frac{\pi}{5}M^5 \nonumber \\
&&+\frac{1}{2}\varphi_c^2( \frac{5\pi}{10}
\frac{\lambda}{\cal{M}}m^3)+\frac{1}{4!} \varphi_c^4 (\frac{45 \pi }{20
}\frac{\lambda^2}{{\cal{M}}^2} m)].
\end{eqnarray}
Thus eqn (19) shows that we have been able to  remove $\Lambda$ successfully and a finite expression for one loop potential using $\varphi^4$ theory in 5D is obtained. Thus it is useful to study effective potential for $\varphi^4$ theory in 5D.
\section{One Loop approximation and Symmetry Restoration}
In this section we  calculate one-loop approximation with temperature corrections. Consider the following Lagrangian
\begin{equation}
\textit{L}= \frac{1}{2} \partial_\mu \varphi  \partial^\mu \varphi -\frac{1}{2} m^2 \varphi^2 - \frac{1}{4!\cal{M}} \lambda \varphi^4 .
\end{equation}
Now shift the field by some constant field $\varphi_c$ and considering quadratic part of shifted Lagrangian, zero loop potential is
\begin{equation}
 V^{0}(\varphi_c)=\frac{1}{2} m^{2} \varphi^{2}_c+\frac{\lambda}{4!\cal{M}}\varphi^{4}_c .
\end{equation}
 From our discussion in previous section one-loop approximation is
given by ($\hbar=1$)
\begin{eqnarray}
V^{1}(\varphi_c)&=&- \frac{i}{2} \int \frac{d^{5}k}{(2\pi)^{5}} \ln(-k^2 + M^2 )\nonumber \\
&=& \frac{1}{2\beta} \sum_n \int \frac{d^{4}k}{(2\pi)^{4}} \ln((\frac{2n\pi}{\beta})^2+\omega^2 ) \\
\omega^2 &=& \textbf{K}^2+M^2 ,
\end{eqnarray}
where $\textbf{K}^2=k^2_1+k^2_2+k^2_3+k^2_4$. From the standard
techniques of four dimensions used previously, we can get a similar
expression for the five dimensions as
\begin{equation}
V^{1}(\varphi_c)= \int \frac{d^{4}k}{(2\pi)^{4}}[\frac{\omega}{2}+\frac{1}{\beta} \ln(1-e^{-\beta \omega})].
\end{equation}
The first term in  eqn (24) is the zero temperature one-loop correction and is  computed in previous section eqn (19). The second term can be written as
\begin{equation}
V^{1}_\beta(\varphi_c)=\frac{1}{8 \pi^{2}\beta^5}  \int^\infty_0 x^3 dx \ln(1-e^{-(x^2+\beta^2 M^2)^\frac{1}{2}}).
\end{equation}
This is the finite temperature contribution to the potential. From this we can calculate $\beta_c$,  following the  method in Ref \cite{3}, i.e using expansion in $V^{1}_\beta(\varphi_c)$
\begin{eqnarray}
V^{1}_\beta(\varphi_c)&=& -\frac{6\zeta[5]}{8\pi^2 \beta^5} +
\frac{2\zeta[3] M^2}{16 \pi^2 \beta^3}
-\frac{M^3}{16\pi \beta^2 } - \frac{3\; M^4}{64 \pi^2
\beta} B   \nonumber \\
&&-\frac{M^4}{32 \pi^2 \beta}\hbox{ln}(M^2\beta^2) + \frac{1}{120 \pi^2}M^5 +
O(M^6 \beta) ~.
\end{eqnarray}
Here, $ B = (\gamma + \frac{\psi_0(\frac{1}{2})}{2}+2-\hbox{ln}(2\pi))\approx -0.2424
$. Now we are in a position to study the symmetry restoration. We
assume the Lagrangian has a symmetry which means that in the absence
of source, $\hat{\varphi}=0$. On the other hand symmetry breaking
occurs when $\hat{\varphi}\neq 0$, indicated by eqn (6). The
complete potential given by eqn (8) can be rewritten into zero
temparature and temperature dependent part as
\begin{equation}
V = V_0(\varphi_c) + V_\beta^1(\varphi_c),
\end{equation}
where $V_0$ is the zero temperature potential including the one loop
correction. The symmetry breaking occurs at zero temperature i.e.
$\frac{\partial V_0(\varphi_c)}{\partial \varphi_c} = 0$ at
$\varphi_c \neq 0$. And the symmetry breaking will disappear when
\cite{3}
\begin{equation}
\frac{\partial^2 V(\varphi_c)}{\partial \varphi^2_c} > 0, \;\;\;
\hbox{for} \;\; \varphi_c \neq 0.
\end{equation}
Now, we see if the temperature dependent part of the potential can
restore the symmetry. Using eqn (27) the symmetry restoration
condition of eqn (28) becomes
\begin{equation}
\frac{\partial^2 V_0(\varphi_c)}{\partial
\varphi^2_c}\vert_{\varphi_c = 0}  + \frac{\partial^2
V^1_\beta(\varphi_c)}{\partial \varphi^2_c}\vert_{\varphi_c = 0}
\geq 0,
\end{equation}
where the first term is the renormalised mass of the theory given by eqn (14).

Thus the critical temperature becomes
\begin{equation}
\frac{\partial^2 V^{1}_\beta(\varphi_c) }{\partial \varphi^2_c} \vert_{\varphi_c =0} = -m^2.
\end{equation}
We obtained an expression for critical temperature as negative of
mass squared and $-m^2$ is considered positive since symmetry
breaking occurs at zero temperature. Thus when $\frac{\partial^2
V^{1}_\beta(\varphi_c) }{\partial \varphi^2_c} \vert_{\varphi_c =0}
> -m^2$, the temperature dependent mass is dominant and the
effective mass becomes positive and condition (28) is satisfied and
hence the symmetry is restored.

Now using eqn (30) we obtain
\begin{equation}
\frac{1}{\beta_c ^3} = \frac{-8 m^2 {\cal{M}} \pi^2}{\zeta[3] \lambda},
\end{equation}
where we  use only on first two terms for calulation of critical temperature  \footnote{One can also consider $\frac{M^4}{\beta}$ term to calculate $\beta_c$, but there will not be significant change in the numerical value of critical temperature.}. 
At temperatures above this critical temperature  the symmetry is restored and below it symmetry will be broken.

Now we will confirm our result, by trying to obtain $\beta_c$  without actually calculating effective potential. That is by calculating self-mass correction
\begin{eqnarray}
m_\beta^2 =m^2 + \delta m^2 + \frac{\lambda}{2 \cal{M}} \int_k \frac{i}{k^2-m^2}.
\end{eqnarray}
Where  $\int_k $ is $\frac{1}{-i\beta}  \sum_n \int \frac{d^4 k}{(2\pi)^4} $,
and hence
\begin{eqnarray}
m_\beta^2&= &m^2 + \delta m^2 + \frac{\lambda}{2\cal{M}} \frac{1}{\beta} \sum_n \int \frac{d^4 k}{(2\pi)^4} \frac{1}{4 \pi^2 n^2/\beta^2 + E_m^2}   \\
&=& m^2 + \delta m^2 + \frac{\lambda}{2\cal{M}}   \int \frac{d^4 k}{(2\pi)^4} \left[ \frac{1}{2 E_m} +  \frac{1}{E_m(e^{\beta E_m}-1)} \right]. \nonumber
\end{eqnarray}
Where we have assumed that mass counter terms cancels with the zero temperature contribution. Thus
\begin{eqnarray}
m^2_\beta &=& m^2 + \frac{\lambda}{16 \pi^2 \beta^3 \cal{M}} \int k^3 dk   \frac{1}{E_m(e^{\beta E_m}-1)}\nonumber \\
&=& m^2 + \frac{\lambda}{16 \pi^2 \beta^3\cal{M}} \int_0^\infty dx \frac{x^3}{(x^2 +\beta^2 m^2)^{1/2}}
\times \frac{1}{(e^{(x^2 +\beta^2 m^2)^{1/2}}-1)} \nonumber \\
&=& m^2 +\frac{\lambda 2 \zeta[3]}{16 \pi^2 {\cal{M}} \beta^3} -
 \frac{3 \; \lambda}{32 \pi {\cal{M}} \beta^2} m +O(\frac{m^2}{\beta}).
\end{eqnarray}
Therefore the critical temperature can be obtained from the following expression
\begin{equation}
0=m^2 + \frac{\lambda}{16 \pi^2 \beta^3_c \cal{M}} \int_0^\infty dx \frac{x^3}{(x^2 +\beta^2_c m^2)^{1/2}} \times
 \frac{1}{e^{(x^2 +\beta^2_c m^2)^{1/2}}-1}.
\end{equation}
It is clear that the value for $\beta_c$ from the eqn (35), relying on $\frac{1}{\beta^3}$ term will be same as in eqn (31). Thus, we obtained the critical temperature for symmetry restoration for the scalar field in 5D.
\begin{center}
\begin{figure}
{ 1 \includegraphics[scale=0.5]{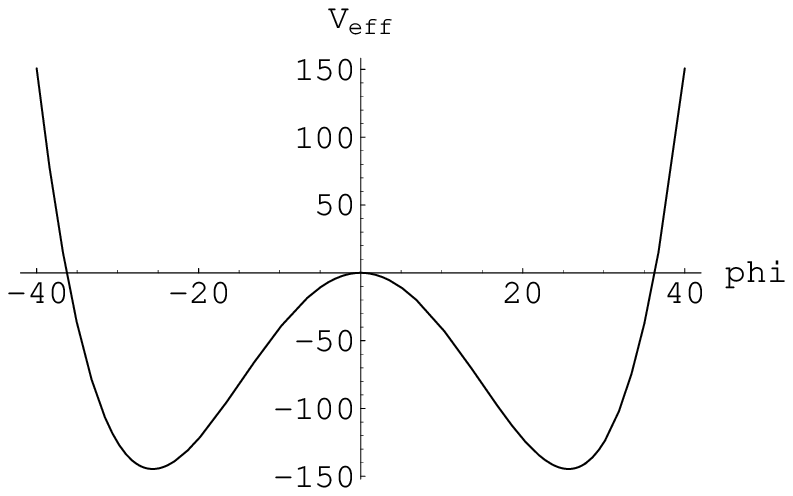}} \hspace{.5cm}
{2\includegraphics[scale=0.5]{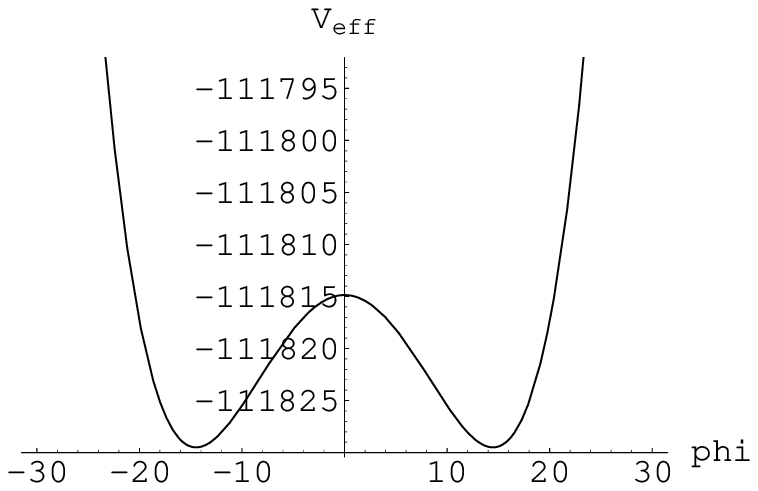}}
\vspace{.5cm}
{3\includegraphics[scale=0.5]{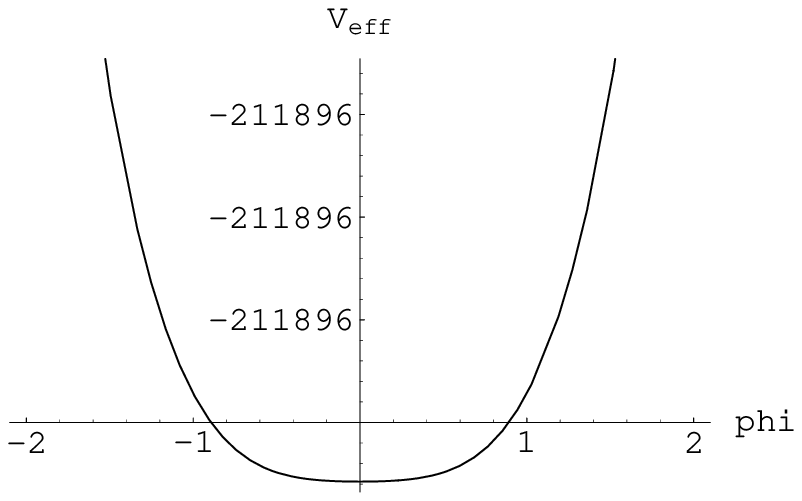}}\hspace{.5cm}
{4\includegraphics [scale=0.5]{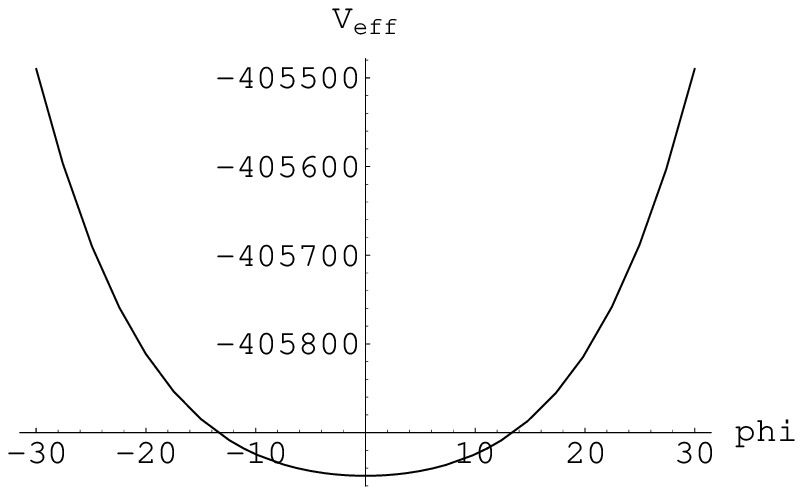}}
\centering{5\includegraphics [scale=0.5]{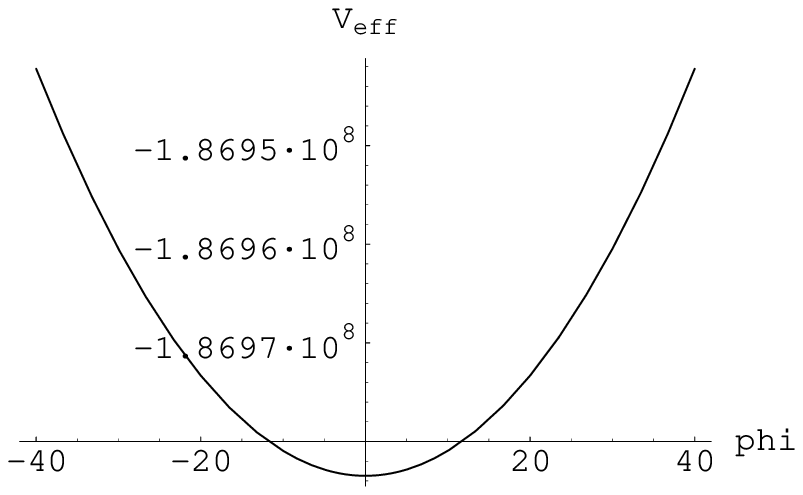}}
\vspace*{8pt}
\caption{Behaviour of finite temperature effective potential  showing symmetry restoration for m = 0.9371, $\lambda=0.008$ and ${\cal{M}}=1$. (1) shows broken symmetry at $T=0$. (2) shows behaviour at $T=17$ less than $T_c$. (3) shows symmetry being restored at $T=19.318$ i.e $T=T_c$. (4) shows behaviour at $T=22$ and symmetry is restored. (5) Shows behaviour of finite temperature effective potential at $T=75$ i.e $T \gg T_c$ where symmetry is completely restored.\label{f1}}
\end{figure}
\end{center}
Next we study the behaviour of symmetry by examining the effective potential ($V_{eff}$). Here we consider lowest order interaction i.e $\lambda\ll 1$, in this limit we  plot $V_{eff}$ as function of $\varphi$ for different values of temperature. Figs 1.1-5 shows behaviour of $V_{eff}$ for different temperatures. At temperature $T=0$ fig(1.1) we have two degenerate minima indicating a symmetry broken phase. The effective potential for $T=17$ which less than $T_c$ still has symmetry broken phase. For $T=T_c$ we observe there are no longer two minima, indicates that it is going to symmetry restored phase. When temperature is more than $T_c$ there is only one minimum at $\varphi=0$ showing restoration of symmetry. At $ T\gg T_c$ effective potential has unique minimum showing symmetry persists. Therefore we observe that the transition from one phase to other phase is continuous and hence it is a second order phase transition.
\section{Discussions}
In this work, we calculated the effective potential for a scalar field up to one loop order. The scalar field is considered in five dimensions and we applied the path integral technique successfully to calculate the effective potential. We are able to regulate $\varphi^4$ theory in 5D and obtain an expression for one loop potential. Using the functional diagrammatic  method,  we calculated temperature dependent one loop correction and the critical temperature $\beta_c$. We note that critical temperature in case of 5D field  is dependent on fundamental mass scale in contrast with 4D case, but still  symmetry restoration is possible, when effective mass squared becomes positive. We verify our result  by  calculating critical temperature from self-mass correction at finite temperature. The nature of phase transition is found  continuous and is  inferred as second order.

The present results can be compared with  Dienes et'al \cite{5} where they consider single extra dimension, which is compactified on a circle and  it contributes a large number of Kaluza-Klein modes. Our expressions for temprature one-loop potential and critical temperature are  similar to there results, in case of a scalar field.
The nature of the phase transition  in this study  is in concordance  with there model \cite{5}.

Our result may find some  applications in higher dimensional cosmological theories,  especially in theories where one would work with higher dimensional fields rather than effective 4D theories where coupling constant would be suppressed due to Planck's scale.

\section*{Acknowledgments}

We would like to thank Prof. A. K. Kapoor, Prof. M. Sivakumar and K. V. S. Shiv chaitanya for useful dicsussions and suggestions during this work.

\end{document}